\documentclass[aps,preprint,unsortedaddress]{revtex4-1}
\usepackage[utf8x]{inputenc}
\usepackage{graphicx}
\usepackage{amsmath, amssymb, setspace}
\usepackage{epstopdf}
\usepackage{color}

\newcommand{\ws}{$2\,\pi\times$28~MHz}
\newcommand{\wy}{$2\,\pi\times$500~MHz}
\newcommand{\wxz}{$2\,\pi\times$400~MHz}
\newcommand{\D}{30~$\mu$m}
\newcommand{\R}{5~$\mu$m}
\newcommand{\alp}{20$^o$}
\newcommand{\Deff}{7.3~$\mu$m}
\newcommand{\depth}{1~meV}
\newcommand{\Vt}{0.4~V}
\newcommand{\Vd}{0.2~V}
\newcommand{\Ad}{350~nm}

\newcommand{\dnh}{8100~motional quanta/s}
\newcommand{\dnl}{690~quanta/s}
\newcommand{\gp}{$2\,\pi\times$1.1~MHz}
\newcommand{\tauswap}{230~ns}
\newcommand{\tauswapd}{560~ns}
\newcommand{\taubelld}{280~ns}
\newcommand{\Glc}{$2\,\pi\times$3~MHz}
\newcommand{\ket}[1]{\left\vert{#1}\right\rangle}
\newcommand{\etfid}{$ 98.8 \rm{\%}\;$}
\newcommand{\etfidbl}{$ 99.4 \rm{\%}\;$}
\newcommand{\eefid}{$ 98.3 \rm{\%}\;$}
\newcommand{\eefidbl}{$ 99.2 \rm{\%}\;$}
\newcommand{\esfid}{$ 99.5 \rm{\%}\;$}
\date{\today}

\begin{document}

\title{Quantum information processing with trapped electrons and superconducting electronics}

\author{Nikos Daniilidis$^1$, Dylan J Gorman$^1$, Lin Tian$^2$, Hartmut H\"affner$^{1,3}$}
\affiliation{$^1$Department of Physics, University of California Berkeley, Berkeley, CA $94720$}
\affiliation{$^2$School of Natural Sciences, University of California Merced, Merced, CA $95343$}
\affiliation{$^3$Materials Sciences Division, Lawrence Berkeley National Laboratory, Berkeley, CA $94720$}

\begin{abstract}
We describe a parametric frequency conversion scheme for trapped charged particles which enables a coherent interface between 
atomic and solid-state quantum systems. The scheme uses geometric non-linearities of the potential of a coupling electrode near a trapped particle. Our 
scheme does not rely on actively driven solid-state devices, and is hence largely immune to noise in such devices. We present a toolbox which can be used 
to build electron-based quantum information processing platforms, as well as quantum interfaces between trapped electrons and superconducting 
electronics.
\end{abstract}

\maketitle

\section{Introduction \label{sec:introduction}}

The experimental realization of an operational quantum computer is a well defined problem \cite{DiVincenzo2000}, but after nearly two decades of intense 
experimental pursuit, the choice of the optimal physical system remains a difficult task \cite{Ladd2010}. Solid-state based systems offer fast 
gate operation times and straightforward fabrication scalability, while atomic systems show remarkable coherence times \cite{Ladd2010,Haeffner2008}.
It appears very appealing to bridge the gap between atomic and solid-state based quantum devices, and combine them into quantum hybrid systems that harness the
benefits of both approaches. Such hybrids can combine the speed of the former with the long coherence times of the latter. Moreover, such platforms 
can interconnect atomic qubits via a solid-state quantum bus \cite{Sorensen2004}, and thus address the scalability challenges of atomic qubits.
Finally, quantum state initialization and read-out can be based on such hybrid interfaces. This is an essential feature for approaches where these tasks are
not straightforward, such as trapped-electron based quantum information processing (QIP) \cite{Ciaramicoli2003}. Here we consider a hybrid system, where 
the long-lived internal state of the atomic system is coherently coupled to an electrical circuit. A successful interface will allow sufficient control 
over the internal degree of freedom of the atomic system, such that we can initialize it in an arbitrary quantum state, swap it with a quantum state in the 
circuit, and read it out with high fidelity. 

In many cases the long-lived state of the atomic system, such as the spin of an isolated particle, couples weakly to electrical circuits, and an 
intermediate system has to be used as a bus \cite{Tian2004,Daniilidis2012}. The motional state of charged particles, for example ions trapped 
in radio frequency (Paul) traps or electrons in Penning traps, can play the role of a bus. Ions can be trapped with very long storage times, 
their motional and electronic quantum state can be controlled to a very high degree, and the motional state can be mapped onto a long-lived 
electronic or spin state of the ion using standard techniques \cite{Haeffner2008}.
A major challenge with ions lies in the frequency mismatch between the ion motion, typically 1 to 10~MHz, and the superconducting circuits, 
with transitions between 4 and 10 GHz. In addition, the small charge induced by the ion motion, of order $10^{-4}$ elementary charges, needs to overcome 
low-frequency noise in the solid state. The frequency gap can be bridged with parametric frequency conversion \cite{Heinzen1990}. One possibility is to actively 
drive some circuit element, for example as proposed by Kielpinski \textit{et al.} \cite{Kielpinski2012} in a scheme achieving upconversion from 1~MHz to 1~GHz. 
This is an exciting option, but time varying contact potentials between macroscopic 
electrodes \cite{Pollack2008} can impact the fidelity of this scheme. Along similar lines, if a parametrically pumped superconducting single electron transistor is used for the frequency 
conversion, $1/f$ charge noise in the SET \cite{Pashkin2009} will be comparable to the ion signal. Thus, there is need of a frequency conversion mechanism which 
upconverts the trapped particle frequency to the microwave range \emph{before} this enters the solid state, since such a scheme would be naturally immune 
to 1/f noise in the solid state.

Electrons offer a number of benefits due to their large charge-to-mass ratio. Most importantly, their motional state has a large electric dipole moment which can 
couple very strongly to electrical circuits \cite{Wineland1975}. They can be trapped with high motional frequencies and long storage times, using 
osicllating trapping potentials in the microwave range \cite{Walz1995,Hoffrogge2011}, or in Penning traps. 
Trapped electron frequencies could reach the microwave regime while operating the electron traps with realistic voltages, provided the trapping structures 
are made sufficiently small, 1 micron or smaller. Nevertheless, such miniaturized traps for electrons are likely to face limitations due to poorly understood 
electric field noise arising from nearby surfaces \cite{Wineland1998, Deslauriers2006a}. Based on measured values of this type of electric field noise at 4~K \cite{Brown2011}, 
a single electron trapped at frequencies of 7~GHz and a 500~nm distance from metallic electrodes kept at cryogenic temperatures, is expected to have energy 
relaxation rate in excess of $10^7$/s, comparable to the coupling rate of 30~MHz between the electron motion and electrical circuits achieved in such a structure 
\cite{Schuster2010}. The electric field noise can be greatly reduced at room temperature \cite{Hite2012}, but it is unknown if the same mechanisms are responsible 
for the noise at cryogenic temperatures, and whether the noise can also be sufficiently reduced in cryogenic conditions. Thus, alternative 
solutions which will work for electrons trapped in larger trap structures, in the several micrometer range, are needed. Pennning traps offer one such possibility 
\cite{Bushev2011}, but complications arise due to the presence of strong magnetic fields if the electrons are trapped in a Penning trap. Thus a frequency 
conversion scheme for electrons trapped in the low magnetic environment of an RF trap would have significant advantages over the above mentioned approaches.

Here we describe a parametric frequency conversion scheme which uses the quadrupolar potential of a trap electrode in combination with classically driven particle 
motion in order to couple the motional degree of freedom of a trapped particle to electrical circuits. This scheme can up-convert the motional signal of any 
charged trapped particle before it enters the solid-state circuit, and thus reduces the impact of $1/f$ noise present in typical solid-state devices. 
We show how, using this scheme, one can cool the motion of a single electron to the ground state, swap the quantum state of an electron and a transmon qubit, 
or entangle the two. Based on these tools, we describe hybrid QIP platforms which are based on single electrons in Paul traps  and superconducting 
microwave electronics. We also discuss the possibility of using this scheme to parametrically couple electrons to circuit elements at higher frequencies, 
above 100~GHz.

The paper is organized as follows: In Sec.~\ref{sec:hamiltonian} we discuss the parametric coupling mechanism, and in Sec.~\ref{sec:trap} we describe one possible ring Paul trap for implementing 
the mechanism with trapped electrons. In Sec.\ref{sec:decoherence} we outline the basic decoherence sources which are expected to limit this scheme. We then 
describe some basic applications of this scheme. These applications form a toolbox which can be used to build several interesting devices. We close with a 
discussion of such possibilities: A QIP platform with electron-spin memory and Josephson-Junction (JJ) processing qubits, and an all-electron QIP platform, 
with JJ qubits used for the electron state readout.

\section{Parametric coupling mechanism \label{sec:hamiltonian}}

The basis of our approach is a parametric frequency conversion mechanism, in which the quadrupolar potential of pick-up 
electrodes near the trapped particle creates the non-linearity which is necessary for the frequency conversion. The pump for the parametric 
process is a classical voltage which drives the particle motion. The non-linear potential of the pick-up electrodes allows us to apply a force 
on the particle which depends both on the signal in an external circuit and on the particle position, which is clasically driven at high 
frequency. This combination gives rise to the parametric action. Here we consider a quadratic non-linearity, i.e. a quadrupolar potential. 

  \begin{figure}[!b]
    \includegraphics[width = 4 cm]{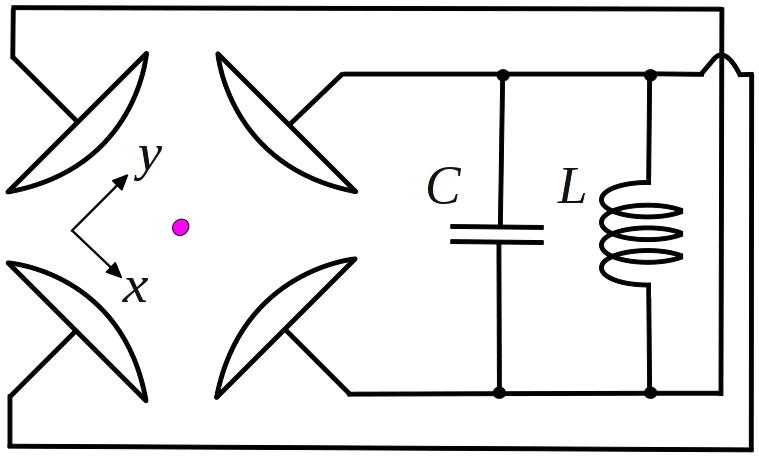}
    \caption{
    Basic setup for parametric frequency conversion. A particle (shown in magneta) is trapped in a harmonic potential, between electrodes 
    which create a strong quadrupole potential around the trapping position. The electrodes are connected to a resonant circuit, allowing the 
    particle motion to couple to the modes of the circuit.
    \label{fig:generic}}
  \end{figure}

We consider a charged particle trapped in a harmonic potential. The particle is located between two sets of coupling electrodes which 
are connected to an electrical resonator, as in Fig.~\ref{fig:generic}(a). The circuit couples to the position of a particle in the trap via the voltage 
on the coupling electrodes. The interaction energy is $qU({\bf r})V$, where $q$ is the charge and ${\bf r}$ the position of the particle, $V$ is the voltage 
between the coupling electrodes, and $U({\bf r})$ the potential at position ${\bf r}$, when 1~V is applied to the coupling electrode. For simplicity, here we 
consider coupling electrodes which create electric quadrupoles of the form $\sum_{i = x,y,z}c_i\,r_i^2$, but the analysis can be generalized to potentials 
containing cross terms as well. For a displacement in the direction $r_i,\, i = {x,y,z}$, around the trapping position, the potential can be expanded as 
$U(r_i) = U(0) + (r_i/D_{1,i})+ (r_i/D_{2,i})^2 + O(r^3)$ \cite{Note}
Then, the Hamiltonian for the trapped particle and the circuit is, to second order in $r_i$
\begin{equation}
 H = \frac{Q^2}{2C}+\frac{\Phi^2}{2L}+\sum_{i}{\frac{p_i^2}{2m}+\frac{m\omega^2 r_i^2}{2}+\frac{e Q}{C}\left(\frac{r_i}{D_{1,i}}+ \frac{r_i^2}{D_{2,i}^2}\right)}\,,
\label{eq:hamiltonian}
\end{equation}
Here, $r_i$ is the particle displamecement, $p_i = m\dot{r}_i$ is the 
particle canonical momentum. $C$ is the effective capacitance of the resonator at the coupling electrode.  $\Phi$ represents the flux variable at the coupling 
electrode, and $Q = C\dot{\Phi}-qU(r_i)+Q_{\rm d}(t)$ is the canonically conjucate charge, which includes the charge, $qU(r_i)$, induced on the electrode by 
the moving particle, and a classical, time-dependent charge $Q_{\rm d}(t)$, induced from the classical parametric drive voltage. The latter is detuned by the 
trap frequency, $\omega_{i}$, from all resonant modes in the system and, as we discuss in Appendix \ref{app:capacitance}, this induced charge can be made 
negligibly small by carefully balancing the different electrode capacitances in the device. 

The coupling term linear in position, $e\,Q\,r_i/(C\,D_{1,i})$, couples the circuit and the particle when the two are resonant, and the quadratic 
terms,  $e\,Q\,r_i^2/(C\,D_{2,i}^2)$, lead to parametric coupling. To switch on the parametric action, we drive classical particle motion,  
$r_{{\rm d},i} = A_{\rm d}\cos(\Omega_{\rm d} t)$, in addition to the quantum motion in the trapping potential, $\hat{r}_{i}$. We decompose 
the particle position as $r_{i} = r_{{\rm d},i}+\hat{r}_{i}$. Expanding the quadrupole part of the interaction energy, we obtain the parametric coupling term
$\frac{2e A_{\rm d}}{C D_{2,i}^2}\cos(\Omega_{\rm d} t)\hat{Q}\hat{r}_i$, where $\hat{Q}$ is the quantum charge degree of freedom in the circuit. When driving 
motion in the $y$ direction, the Hamiltonian in the interaction picture now becomes 
\begin{equation}
 H_{\rm er} = \hbar g \cos(\Omega_{\rm d} t) \left( e^{i (\Omega-\omega_y)t}a_{\phi}^\dagger a_{y}+ e^{i (\Omega + \omega_y)t}a_{\phi}^\dagger a_{y}^\dagger + h.c.\right)\,.
 \label{eq:hamiltonian-int}
\end{equation}
The $a_{\phi}$, and $a_{y}$ operators correspond to the circuit and particle modes respectively, $\Omega = 1/\sqrt{LC}$ is the circuit 
resonant frequency,  $\omega_y$ the particle frequency, $ \hbar g = \frac{2 e q_0 A_{\rm d} y_{0} }{C D_{2,y}^2}$,
 $y_{0} = \sqrt{\hbar/(2 m \omega_{y})}$ describes quantum fluctuations of the particle position, and $q_0 = \sqrt{\hbar/(2Z)}$ 
quantum fluctuations of the circuit charge variable, which depends on the characteristic impedance $Z = \sqrt{L/C}$. 

If $\Omega_{\rm d} = \Omega-\omega_{y}$, then the terms $\left(i\,a_{\phi}^\dagger a_{y} + h.c.\right)$ 
of Eq.~\ref{eq:hamiltonian-int} survive in the rotating wave approximation . 
The system operates as a parametric frequency converter, with the classical drive providing 
pump photons which allow coherent coupling between the particle and the resonator. Population exchange between the 
two modes occurrs with a parametric coupling rate $ g_{\rm p} = g/2$ \cite{Louisell1961}. If $\Omega_{\rm d} = \Omega+\omega_{y}$, then 
the system behaves as a parametric amplifier \cite{Louisell1961}. The effective Hamiltonian
then has the form $\left(i\,a_{\phi}^\dagger a_{y}^\dagger + h.c.\right)$ which generates two-mode squeezing of the coupled modes \cite{Braunstein2005}. 

In what follows, we focus on the former of these two cases. We also focus on electrons, which due to their large charge-to-mass ratio can couple very strongly 
to microwave circuits using currently attainable experimental parameters. In Sec.\,\ref{sec:applications} we discuss applications of this scheme: 
i.e. quantum state initialization for the electron, creation of entanglement and quantum state transfer between single electrons and superconducting 
qubits, as well as creation of entanglement and quantum state transfer between distant elecrons.

\section{Electron trap and resonator design \label{sec:trap}}

\begin{figure}[!b]
    \includegraphics[width = 8 cm]{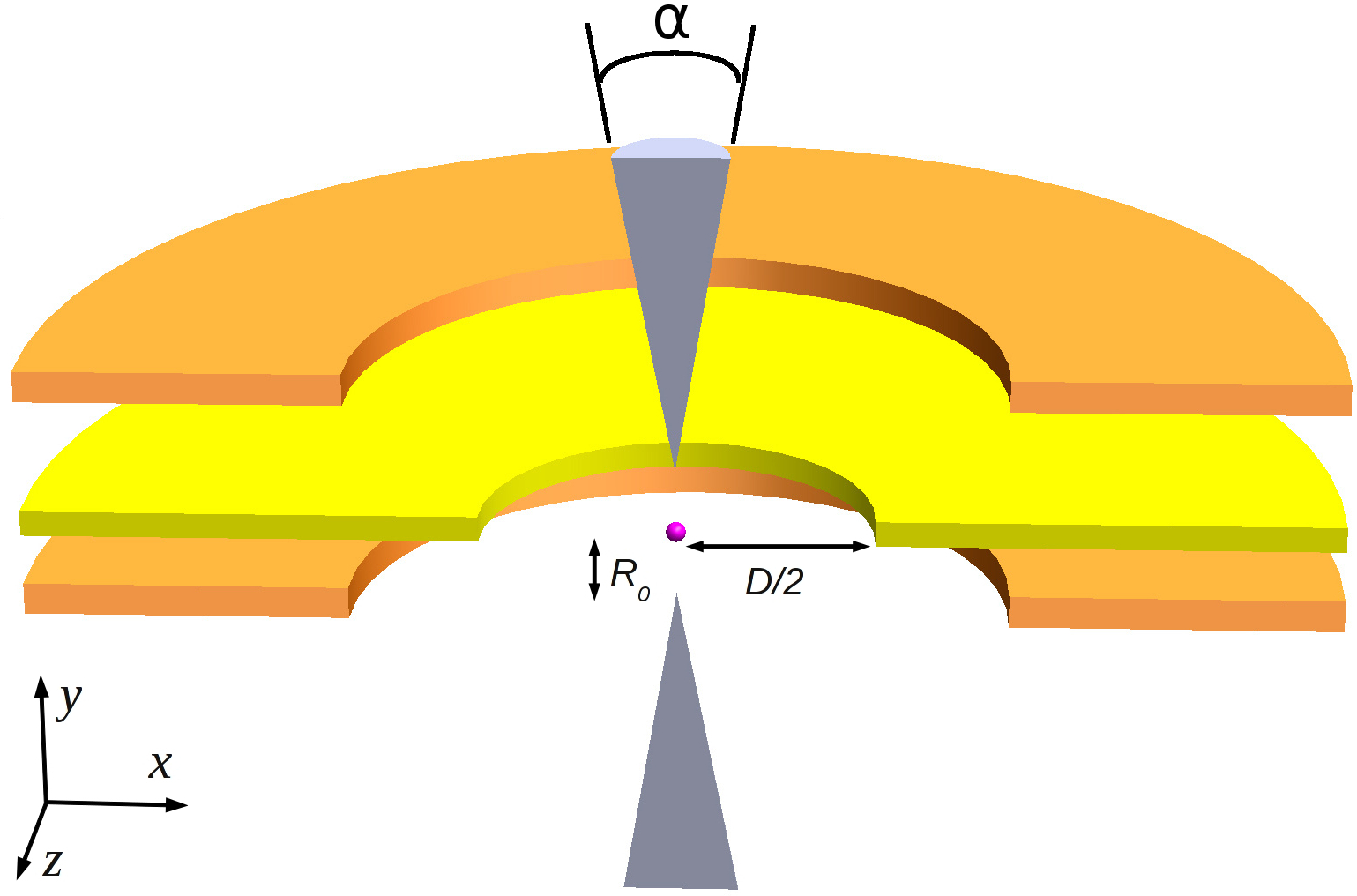}
    \caption{
    A ring trap with two sharp tip electrodes (trapped electron in magenta). The trapping ponderomotive pseudopotential is created by a ring 
    electrode (yellow) with inner diameter $D$. Two conical tip electrodes (grey) with opening angle $\alpha$ are located a distance $R_0$ from the 
    trapping position in the $y$ direction. These can be connected to an external circuit, allowing to couple the particle motion 
    to the circuit. Two ring electrodes (orange) can classically drive the particle motion. To achieve parametric coupling, we drive electron motion 
    in the $y$ direction with amplitude $A_{\rm d}$, but driving motion in the $xz$ plane is also possible. An equivalent alternative configuration 
    for trapping and parametric coupling, is to connect the middle ring electrode (yellow) to an electrical resonator, connect the top and bottom 
    tip electrodes to the source of the trapping ponderomotive potential, and use the top and bottom ring electrodes (orange) for the parametric drive.
    \label{fig:trap}}
  \end{figure}

For simplicity, we choose a ring trap to trap single electrons  (Fig.~\ref{fig:trap}). This kind of trap combines high trap depth, low anharmonicity of the 
trapping potential, and strong parametric coupling. We simulated this design with  $D = $\D, $R_0 = $\R, $\alpha =$\alp  ~ (see Fig.~\ref{fig:trap} for 
an explanation of the parameters) using an 
electrostatics solver \cite{Singer2010}. The effective coupling length appearing in Eq.~\ref{eq:hamiltonian} is $D_{\rm 2,y} = $ \Deff. 
Single electrons with secular frequencies $\omega_{y} = $\wy, $\omega_{x,z} \approx $\wxz, 
can be trapped with trap depth of \depth~using a trap drive on the central ring electrode (shown in yellow) at 
$\Omega_{\rm tr} \approx 2\pi\times7$~GHz, amplitude approximately \Vt, and with a static bias of a few hundred mV on the trap electrodes. 
To implement the parametric coupling scheme, we can drive electron motion in the $y$ direction at $\Omega_{\rm d} =\Omega_{\rm tr}$ and $A_{\rm d} = $\Ad, 
by applying opposite oscillating voltages of amplitude \Vd~on the top and bottom ring electrodes (orange). Numerical integration of the equations of motion 
shows that the trap is stable under this condition (see Appendix \ref{app:param-drive}). The trapping potential and the parametric pump drive will not significantly limit the fidelities of 
processes described in Sec.~\ref{sec:applications}, if they are stable to better than 1 part in $10^{3}$. The capacitances between the tip electrodes 
and the ring electrodes in this structure range from 0.3~fF to 0.8~fF. While this will have only a small loading influence on the resonator to which the 
particle motion will couple, the resonator can be off-resonantly excited by the parametric drive and the trapping potential. We discuss solutions to 
these technical issues in Appendix \ref{app:capacitance}.

To load single electrons in the ring trap, one option is to have the trap fabricated at the end of a linear Paul trap with segmented 
electrodes \cite{Kielpinski2002}. The linear trap can have a taper from large trap dimensions to smaller dimensions \cite{Amini2010} to load electrons at high 
energy and resistively cool them \cite{Wineland1975} in different stages (e.g. precooling to 10 K, followed by cooling to 1 K to load into the ring trap). Electron
clouds can be loaded 
in the linear trap using a heated filament, or, in order to have better control on the number of created electrons, by photoionization of an atomic vapor. 
After the electron cloud is cooled to 1 K, the number of electrons in the trap can be distinguished by coupling their motion to an electrical resonator 
at the electron resonance frequency \cite {Wineland1975}, and the segmented trap electrodes can be used to heat and split the electron cloud until a single 
electron is trapped \cite{Wineland1973}. Finally, the electron can be transported into the ring trap, and `locked' in place by modifying the ponderomotive 
trapping potentials of the linear trap and the segmented trap \cite{Karin2011,Kumph2011}.

The resonator depicted schematically in Fig.~\ref{fig:generic} can be a lumped-element resonator, or a coplanar waveguide (CPW) resonator. The coupling strength 
between an electrical resonator and the particle in the trap will benefit from high characteristic impedance resonators, due to the $\sqrt{Z}$ dependence of 
quantum voltage fluctuations on the characteristic impedance $Z$. The effective impedance $Z$ for a CPW section with length $n\lambda/4,\,n=1,2,...$ is 
related to the CPW characteristic impedance, $Z_{\rm CPW}$, by $Z = \frac{4\,Z_{\rm CPW}}{n\,\pi}$ \cite{Daniilidis2012}. In what follows, we consider a 
resonator with characteristic impedance 1~k$\Omega$. TiN-based high kinetic inductance resonators \cite{Zmuidzinas2012} are promising in this respect.  
Using this technology, resonators with very high inductance per unit length, exceeding $\approx60\,{\rm pH}/{\rm \mu m}$, have been achieved \cite{DayPrivComm}. 
Designing resonators based on such films, with gap between the center conductor and the ground plane in the tens of ${\rm \mu m}$ range would achieve the 
required impedance of approximately $ 1 {\rm k \Omega}$.

\section{Decoherence sources\label{sec:decoherence}}

Provided a sufficiently low-noise classical drive, parametric frequency conversion can couple two non-resonant systems with no added noise \cite{Mishkin1969}. 
The fidelity of coupling between electrons and electrical circuits will be limited by motional decoherence of the electron motion, decoherence in the 
resonator and superconducting qubit circuits, and classical noise in the trap drive and the parametric drive. 

To estimate the heating rate of the electron motion in the $y$ direction, we need to know the spectral density of electric field noise at $\omega_{y}\approx$\wy \cite{Deslauriers2006a}. 
Johnson noise and electronic technical noise can be made very small, so we focus on the so called 'anomalous' heating, encountered in ion traps. The dominant 
contribution  of this noise has been shown to arise from the electrode surfaces \cite{Hite2012}. We can model the noise as arising from a collection of 
independently fluctuating electrical-dipole type sources on the trap electrodes, in which case the noise level is determined by the surface density of electrical dipoles 
on the electrodes \cite{Daniilidis2011,Safavi-Naini2011}. In this model, the magnitude of the noise for a given density of dipoles has been shown to depend on the 
electrode geometry \cite{Low2011}. We take into account the non-planar geometry of the proposed trap as discussed in Appendix \ref{app:el-heating}. 
Based on heating rates measured in ion traps at 4\,K \cite{Brown2011}, a single electron trapped at $\omega_{y}$ = \wy~will have $\tau_{1}\approx123 \,\mu$s
(heating of \dnh) if the noise scales with frequency as $f^{-1}$, and $\tau_{1}\approx1450 \,\mu$s (heating at \dnl) if the scaling is $f^{-3/2}$, as observed in \cite{Hite2012}.

The internal quality factor ($Q_{\rm i}$) of CPW resonators is thought to be limited by fluctuating two-level systems (TLS) in 
the interface between the superconductor and the dielectric substrate on which it is fabricated \cite{Gao2007,Gao2008,Kumar2008}. As a result, $Q_{\rm i}$ 
decreases by 1 to 2 orders of magnitude as the the energy stored in the resonator decreases to the few photon level. In recent years, significant efforts in 
dielectric substrate cleaning and materials engineering have resulted in an increase of $Q_{\rm i}$ \cite{Wang2009,Vissers2010}, with values at the single photon 
level currently exceeding $10^6$ \cite{Megrant2012}. Moreover, it has been realized that the resonator losses can be limited by reducing the participation of the 
dielectric-superconductor interface in the resonant mode. One way to achieve this is by building higher characteristic impedance CPW resonators 
\cite{Geerlings2012}. This can prove advantageous for the high characteristic impedance resonators $Z_{\rm CPW}\sim1\,{\rm k}\Omega$ needed in our application. 
TiN-based high kinetic inductance resonators in the $2\,\pi\times$1-2~GHz range, already mentioned in Sec.~\ref{sec:trap}, show very high quality factors \cite{LeDuc2010a}, and due to their high kinetic 
inductance have wavelength significantly lower than the vacuum wavelength, which significantly reduces their radiative losses. In what follows, we assume a 
resonator with quality factor similar to the best value obtained by Megrant \emph{et al.}, with $\tau_{1}=45\,\mu$s at $\approx2\,\pi\times$7~GHz \cite{Megrant2012}. 

The main goal of this work is to show how to couple an electron to a superconducting qubit, via the above mentioned resonator. We now discuss the superconducting 
qubit which currently exhibits the best coherence times, namely the `3-dimensional' transmon qubit \cite{Paik2011,Rigetti2012}. 
The transmon is a `Cooper pair box' qubit in which the Josephson junction capacitance is increased to make the device largely immune to charge noise \cite{Koch2007}. 
Recently, this kind of device has shown coherence times in the several tens of $\mu {\rm s}$ range by suppressing radiative and charge-flucutator related losses 
after placing the device inside microwave cavities \cite{Paik2011,Rigetti2012}. Here we focus on this implementation of superconducting qubits, and assume decoherence 
times $\tau_{1} = 70\, \mu{\rm s}$ and $\tau_{2} = 92\, \mu{\rm s}$, as those in \cite{Rigetti2012}. 

\section{Applications \label{sec:applications}}

\subsection{Electron-resonator coupling}

\begin{figure}
    \includegraphics[width = 2.85 cm]{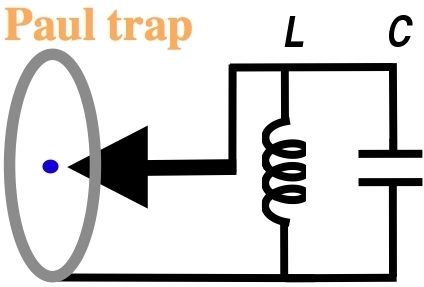}
    {\small (a)}
    \includegraphics[width = 5.5 cm]{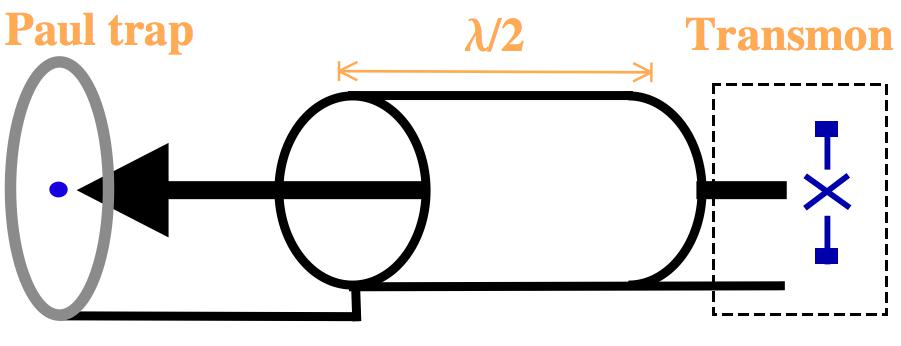}
    {\small (b)}
    \caption{Circuits for electron-resonator and electron-transmon coupling. 
    (a) An electron in a ring trap, with the tip electrodes connected to a microwave LC resonator  with impedance $Z = \sqrt{L/C}$. 
    (b) Schematic drawing of an electron coupled via a $\lambda/2$ section of a coplanar transmission line to a transmon in a cavity \cite{Paik2011}. 
    For superconducting coplanar resonators in the GHz range, internal quality factors of more than $10^6$, corresponding to damping times of $\approx$45~$\mu$s, 
    have been achieved \cite{Megrant2012}. To achieve high characteristic impedance, $Z = 1\,{\rm k}\Omega$, and internal quality factors $Q>10^6$, 
    high kinetic inductance resonators based on thin TiN films can be used \cite{LeDuc2010a}.
    \label{fig:circuit}}
  \end{figure}

In order to couple the electron to a microwave circuit, we consider the tip electrodes to be connected to the open end of a $\lambda/4$ superconducting 
coplanar waveguide resonator (CPW), or a lumped element resonator (Fig.~\ref{fig:circuit}a). In the case of a CPW, quantization of the resonator mode can be 
treated as in \cite{Blais2004}. 
With trap frequency $\omega_{y} = $\wy, driven motion $A_{\rm d} = $\Ad, $\Omega = 2\pi\times$7~GHz, and $Z = 1\,{\rm k}\Omega$,
the coupling rate is $g_{\rm p} = $\gp. This allows complete population exchange between the motion of 
a single electron and a $2\pi\times7$~GHz resonator in $\tau_{\rm swap}\approx$\tauswap. 
By turning on the parametric coupling between an electron resistively precooled to $\sim1\,$K \cite{Heinzen1990} and a microwave resonator at 30\,mK, for 
time $\tau_{\rm swap}$, the electron motion can be prepared to its ground state with approximately 99.8\% fidelity. 
The fidelity of this operation is limited by the heating of the electron motion during the swap operation, and can serve as a quantum-state initialization step in the 
context of QIP. 

\subsection{Electron-transmon coupling}

For a specific example of a hybrid quantum device realizable under our scheme, we consider the case of coupling an electron to a transmon 
through an intermediary transmission line, as shown schematically in Fig.~\ref{fig:circuit}. The tips of the Paul trap are connected to 
the open end of a $\lambda/2$ CPW resonator, which couples the $y$ electron oscillation to the resonator. The transmon is operated inside a 
3 dimensional cavity, an architecture which provides increased coherence 
times \cite{Paik2011}. The second open end of the $\lambda/2$ resonator extends into the cavity, allowing it to couple to the 
${\rm TE}_{011}$ mode of the cavity with a rate $G_{\rm lc}\approx$\Glc~(Appendix \ref{app:cpw-cavity}). The transmon is very strongly 
coupled to the cavity, with coupling constant $G_{\rm tc}$ in the $2\pi\times$100~MHz regime \cite{Paik2011}. The cavity-transmon detuning satisfies 
$\Delta = \Omega_{\rm c}-\Omega_{\rm t}\gg G_{\rm tc}$, i.e. the system is operated in the dispersive regime and the state that the 
resonator couples to is a dressed transmon state with transition frequency $\omega_{\rm t}$. Adiabatically eliminating the cavity, yields an 
effective coupling rate $G_{\rm{lt}} = G_{\rm{lc}}G_{\rm{tc}}/\Delta$ between the transmission line and the dressed transmon (Appendix \ref{app:dressed-transmon}).  
The effective Hamiltonian for the electron-resonator-transmon system is:
 \begin{equation}
  H_{\rm et} = H_{\rm er} + \hbar g_{\rm p} \left( e^{- i \delta t} a_{\phi}^\dagger \sigma^-  + h.c  \right) \,,
 \label{eq:hamiltonian-et}
\end{equation}
where $\sigma^-$ is the Pauli spin lowering operator for the transmon qubit, we have allowed for a detuning 
$\delta = \Omega_{\rm r}-\omega_{\rm t}$ between the resonator and the transmon, and we choose a 
parametric drive $\Omega_{\rm d} = \omega_t -\omega_{y} - \delta$ in $H_{\rm er}$ (Eq.~\ref{eq:hamiltonian-int}). 
To optimize state transfer we choose $\Delta$ such that $G_{\rm{lt}} = g_{\rm p}$.

The detuning $\delta$ is necessary to produce maximally entangled states of the electron motion and the transmon (i.e. Bell states), 
and reduces the decoherence induced 
by losses in the bus. For an arbitrary detuning, this Hamiltonian will not generate complete state transfer between the electron 
and the transmon, because some population will, in general, be left in the transmission line. However, by choosing a ``magic" detuning 
$\delta_n = \sqrt{\frac{8n^2}{2n+1}}g_{\rm p},\, n = 1,2,\ldots$ full state exchange will occur between the electron and the transmon in 
$\tau_{\rm swap} = \frac{\pi}{g_{\rm p}}\sqrt{\frac{2n+1}{2}}$, and the two in a Bell state at $\tau_{\rm swap}/2$. 
This situation is similar to the M{{\o}lmer S{\o}rensen gate for trapped ions \cite{Sorensen2000b}. Using the parameters quoted above for the electron 
traps and for the microwave resonator, electron-transmon state transfer is achieved in \tauswapd. By numerically solving the Lindblad master 
equation of the coupled system (see Fig.~\ref{fig:populations}), we find a fidelity for state exchange of \etfid for the $n = 1$ magic detuning. 
At time $\tau_{\rm swap}/2\approx$\taubelld~the electron and the transmon are in the Bell state 
$\frac{1}{\sqrt{2}}\left(\ket{0,1}-\rm{i}\ket{1,0}\right)$ with fidelity \etfidbl. With our set of parameters, these fidelities are limited mainly 
by losses in the bus and by heating of the electron motion. For the $n=0$ magic detuning, an electron-transmon swap operation 
is completed in 320~ns with fidelity 99.4\%.

\subsection{Electron-electron coupling}

An additional application of this parametric scheme is in coupling electrons in separate traps via a microwave bus. If both ends of the 
$\lambda/2$ CPW are connected to the coupling tips of two electron traps, the electron in each trap gets coupled to the microwave bus with parametric 
coupling constant $g_{\rm p}$. Using the same 'magic' detuning idea as above and the parameters of Fig.~\ref{fig:populations}, we find that the two 
motional states can be entangled with each other within $\tau_{\rm swap}/2\approx$\taubelld~with fidelity \eefidbl, and swapped within 
$\tau_{\rm swap}\approx$\tauswapd, with fidelity \eefid. For the $n=0$ magic detuning, an electron-electron swap operation 
is completed in 320~ns with fidelity 99.1\%.

  \begin{figure}
     \includegraphics[width = 8 cm]{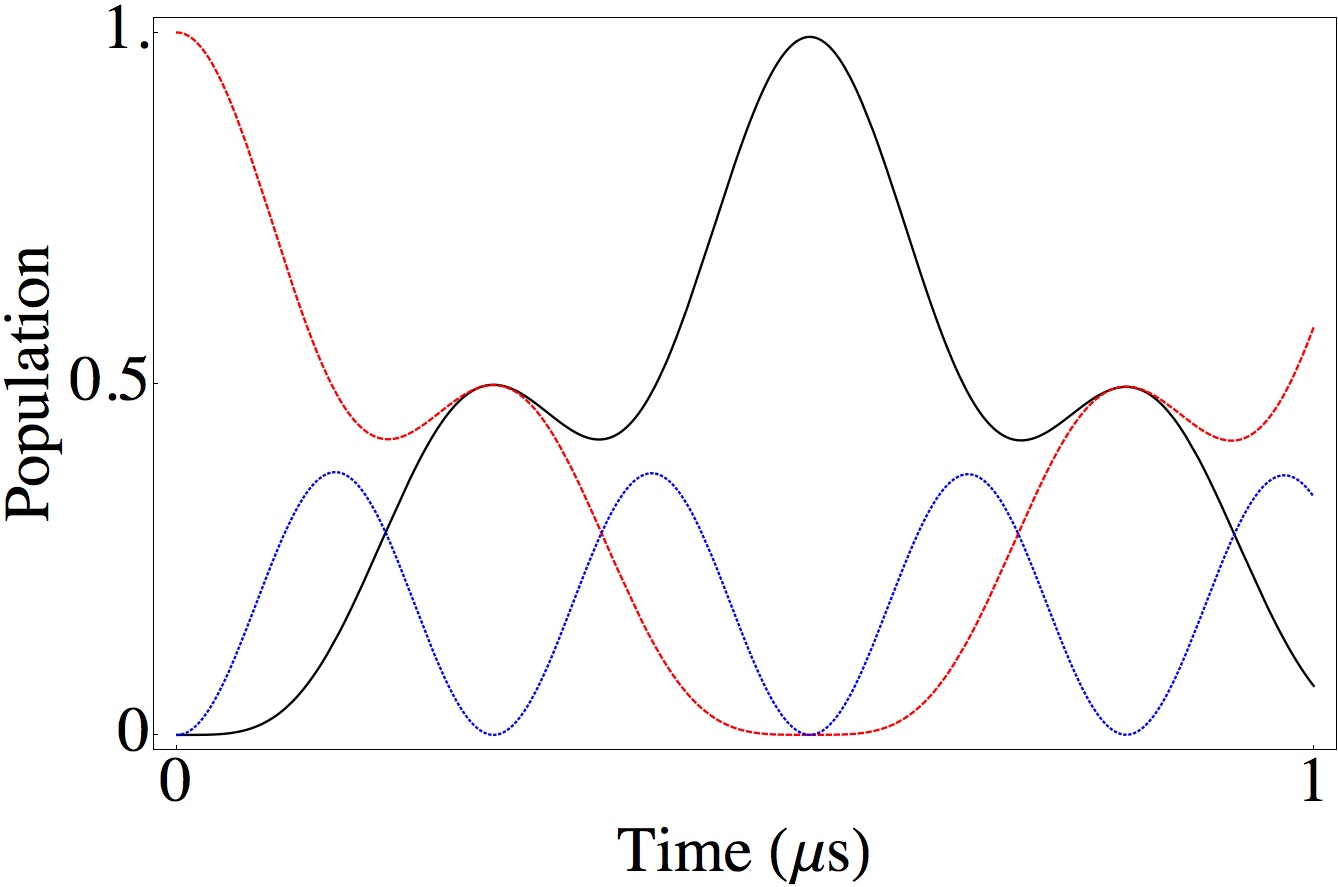}\\
      \vspace{0.3cm}
   \caption{
    Populations derived from a numerical solution of the Lindblad master equation for the electron motional mode ($\left<a_{y}^{\dagger}a_{y}\right>$, red), 
    the transmission line ($\left<a_{\phi}^{\dagger}a_{\phi}\right>$, blue), and the dressed transmon ($\left<\sigma_{z}\right>$, black), for the $n = 1$ magic detuning 
    (see text).
    The parametric coupling rate is $g_{p} = $\gp.
    The initial state is $\ket{1,0,0}$, and the fidelity of evolution to the state $\ket{0,0,1}$ after $\tau_{\rm swap} \approx$\tauswapd~is \etfid. 
    At $\tau_{\rm swap}/2$, the electron and the transmon are entangled in the state $\ket{1,0}-\rm{i}\ket{0,1}$, with fidelity \etfidbl. For the $n=0$
    detuning, an electron-transmon swap operation is completed in 320~ns with fidelity 99.4\%, and an electron-electron swap with fidelity  99.1\%.
    The electron heating rate is \dnh~($\tau_{1}\approx123 \,\mu$s), the transmon decoherence times are $\tau_{1} = 70\,\mu$s, 
    $\tau_{2} = 92\,\mu$s \cite{Rigetti2012}, and the resonator damping time $\tau_{1} = 45\,\mu$s \cite{Megrant2012}.
    \label{fig:populations}}
  \end{figure}

\subsection{Spin-motion coupling \label{sec:spin}}

In order to take full advantage of the low decoherence of the trapped electron system, we now consider mapping the electron motional state to its 
spin. We can define an electron spin manifold with splitting in the radio-frequency range, e.g. $\omega_{\rm s} = 2\pi\times$28~MHz using a static 
bias field of $10^{-3}$~T, see Fig.~\ref{fig:architectures}(a). To do the state mapping, 
we consider the coupling mechanism implemented already with trapped ions \cite{Ospelkaus2008a,Ospelkaus2011}. Microfabricated coils near the trap 
generate an oscillating magnetic field with frequency $\omega_{y}-\omega_{\rm s}$, thus driving a transition between the electron motion and 
its spin. Using a Helmholtz coil geometry with radius $50\,\mu$m, driven such that only a quadrupole magnetic field is generated at the electron, 
an oscillating current of 1~A, and frequency of 272~MHz can drive 
spin-motion transitions with Rabi frequency $2\pi\times410$~kHz. Here, we assumed again $\omega_{y}=$\wy, and 
$\omega_{\rm s} = 2\pi\times$28~MHz, corresponding to a static bias field of $10^{-3}$~T. 
The electron motional state can be mapped onto the spin in approximately 610~ns, with \esfid~fidelity. The coils which generate the oscillating 
magnetic fields can be thermally anchored on a 1~K refrigeration stage to minimize heat load on the 30~mK stage, which is necessary for the 
superconducting electronics.

In order to preserve the phase coherence of the electron spin, the magnetic field at the electron needs to be stabilized. By stabilizing 
the magnetic field to 14~$\rm{pT}/\sqrt{{\rm Hz}}$, the coherence time of the electron spin will exceed 1~s. This noise requirement is 
rather modest, it is three orders of magnitude less stringent than those achieved with magnetic field shielding in SQUID magnetometry \cite{Robbes2006}. 
Heating of the electron motion in a spatially inhomogeneous magnetic field will cause additional dephasing. This can be mitigated by 
engineering a homogeneous static magnetic field, and by periodically cooling the electron motion to its ground state. 

\section{Outlook \label{sec:outlook}}

The elementary toolbox described above can be used in hybrid QIP platforms in which the electron spin serves as a quantum memory, and the 
electron motion as a bus for coupling to superconducting circuits, see Fig.~\ref{fig:architectures}. One possibility is for the transmon qubits to function 
as processing units, and the electron spins to serve as a quantum memory (Fig.~\ref{fig:architectures}b). A second possibility is to use the trapped electrons 
as both processing and memory units, with the transmon serving as a state readout device. A third option uses moving electron qubits in segmented linear Paul 
traps, much the same way in which ion-trap based scalable QIP is pursued (Fig.~\ref{fig:architectures}c) \cite{Kielpinski2002,Hanneke2009a}. 

The first two types of architecture can be implemented using the building blocks shown in Fig.~\ref{fig:architectures}(b). In both cases, 
the LC resonator-based ground state cooling of the electron, and the magnetic-field based spin-motion coupling serve to initialize the electron state. 
If the superconducting qubits are used for information processing, the ${\rm SWAP}$ operation between electron and transmon allows information exchange between 
the processing and memory qubits. In the case of electron-based QIP, ${\rm SWAP}$ operations allow transfer of information between different nodes. 
Single qubit rotations can be performed on the electron spin, which together with the $\sqrt{\rm SWAP}$ gates between the motion of different electons offer 
a universal set of gates. One way to read out the state of the electron is by coupling the electron motion to a dressed three-dimensional transmon, as described 
above, but alternative architectures would be sufficient for this task. In fact, coherently swapping the electron motion with the field of a microwave 
resonator opens the possibility of coupling the electron to any type of superconducting non-linear device which can be coupled to microwave resonators. 

  \begin{figure}
    \includegraphics[width = 8 cm]{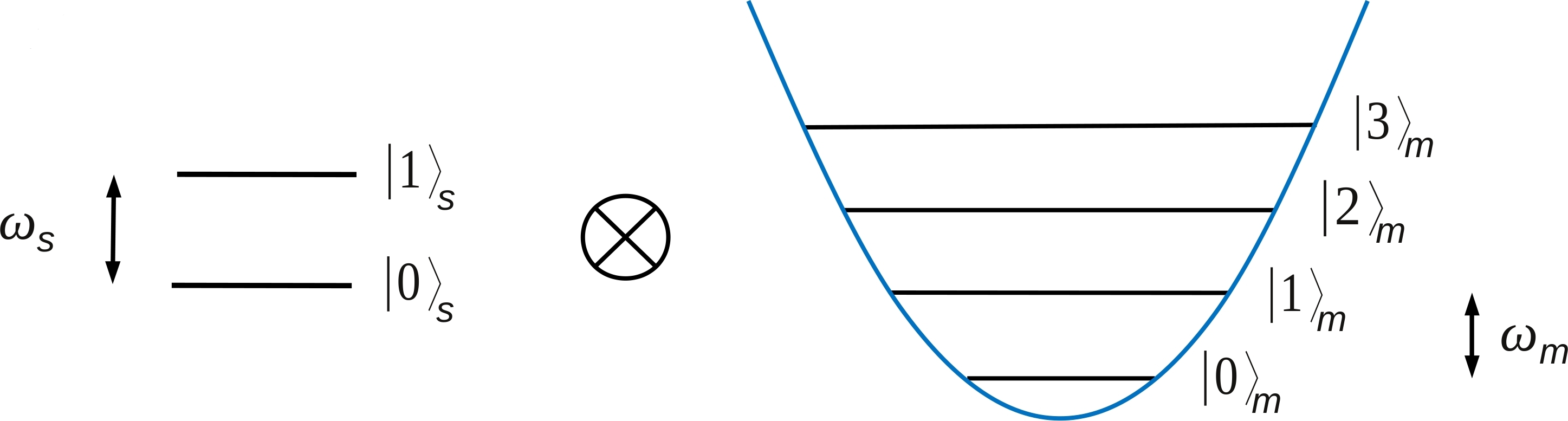}
    {\small(a)}
    \includegraphics[width = 6 cm]{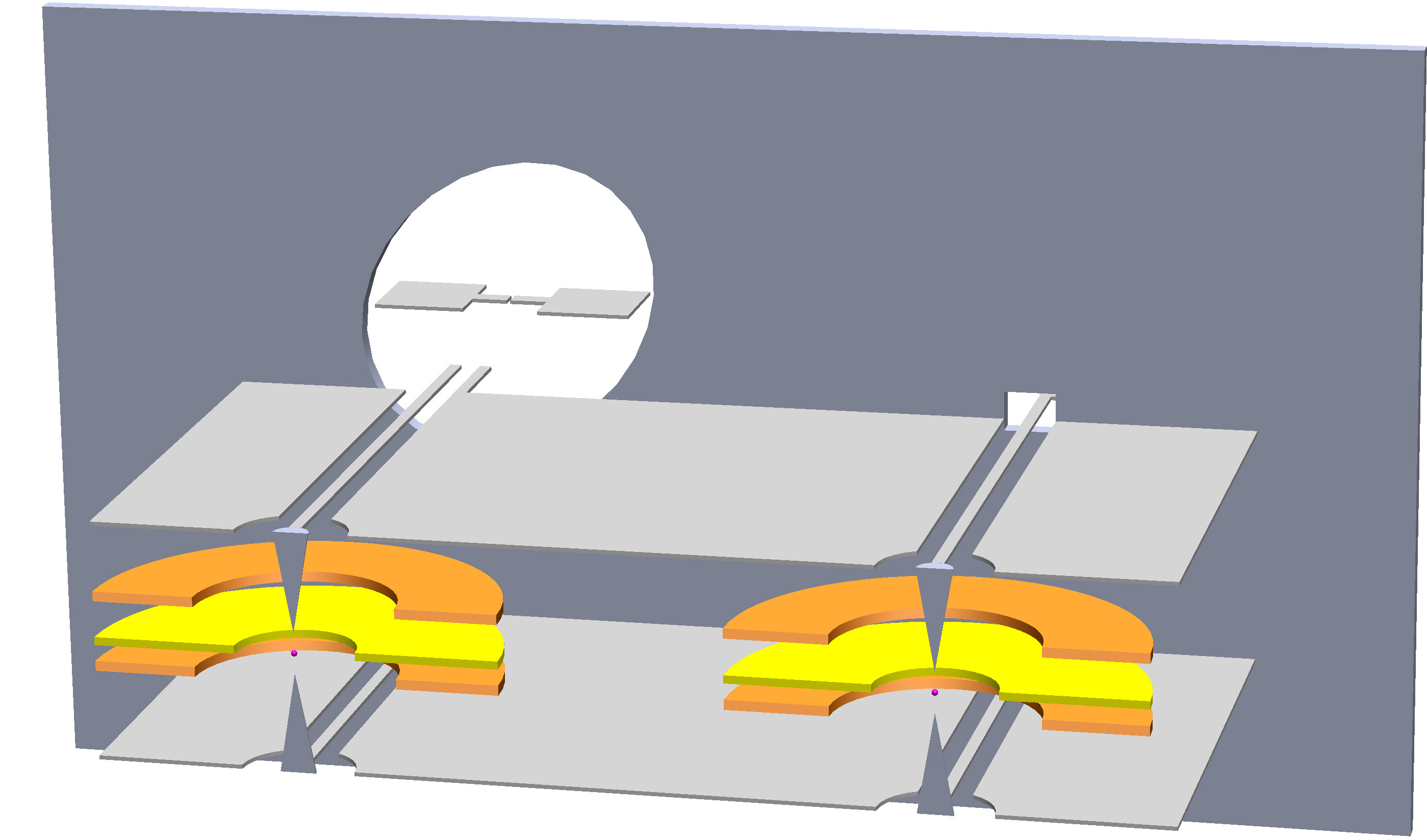}
    {\small (b)}\\
    \vspace*{0.5cm}
    \includegraphics[width = 6 cm]{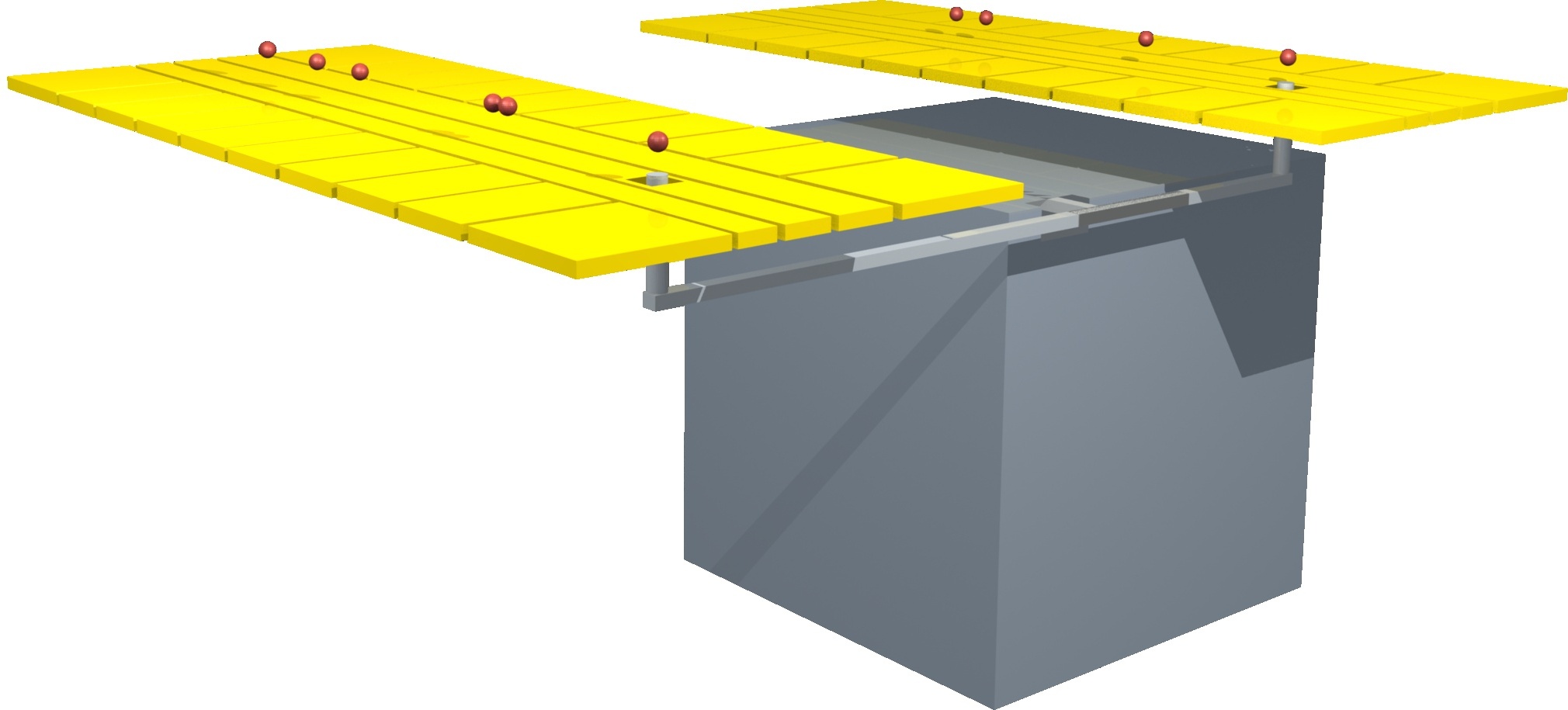}
    {\small (c)}
    \caption{Single electron qubit, and devices using the toolbox developed here.
    (a) Our scheme uses the electron spin as a quantum memory, and one mode of its harmonic motion in the trap as a bus for 
    coupling to electrical circuits. A static magnetic field provides a splitting $\omega_{s}$ of the electron spin manifold, 
    defining the two-level system used to store quantum information. Under a magnetic field of $10^{-3}$~T, the electron Zeeman splitting is 
    $\omega_{s} = $\ws. Typical electron frequencies for the applications we describe will be approximately $\omega_{x} = \omega_{z} = $\wxz, $\omega_{y} = $\wy. 
    (b) Schematic of an electron-transmon hybrid. Transmons operating inside 3d cavities are the processing qubits, and each 
    one is coupled to an electron quantum memory via a $\lambda/2$ resonator. 
    (c) Schematic of an all-electron architecture. Electrons are trapped on a segmented Paul trap. They can be shuttled to regions where their motion is parametrically 
    coupled to microwave resonators and to transmon qubits (inside grey box). Electron-electron gates can be performed via direct Coulomb interaction for electrons on 
    the same trap chip, and using a microwave bus for electrons on different chips. 
    In (b) and (c), the electron traps can be operated at a different temperature stage (e.g. 1~K, yellow \& orange) from the superconducting electronics
    (30~mK, grey) to minimize the heat load on the latter. 
    \label{fig:architectures}}
  \end{figure}

The third distinct architecture which becomes possible using the toolbox described here uses moving electron qubits. Our proposed parametric frequency 
conversion mechanism can be applied to linear microfabricated Paul traps for electrons, similar to the ones extensively pursued for trapped ions \cite{Hanneke2009a}. 
In this case, ground state cooling of the electron motion, state initialization, and readout of the electron spins can be based on microwave circuits. 
Two-qubit entangling gates can be performed using a microwave bus, the direct Coulomb interaction between nearby electrons \cite{Ciaramicoli2003}, or 
with microwave gates \cite{Ospelkaus2008a}. Finally, non-linear superconducting circuits can be used to read out the state of the electrons. 
This approach will not require lasers for cooling, manipulating, and detecting the electron qubits, as trapped-ion based approaches do. In addition, 
it can be significantly faster than current ion-trap based approaches. State initialization and read-out can be performed on the order of a few $\mu$s,
roughly two orders of magnitude faster than with ions. Owing to the higher electron frequencies, particle transport can also be two orders of magnitude faster.
Two-electron gates based on the Coulomb interaction of nearby electrons, will be limited by the rate of spin-motion coupling that can be achieved. This can
be more than one order of magnitude faster than the values achieved with ions, due to the larger extent of the electron's wavefunction. 

As a final, longer-term application, we consider the possibility of scaling an architecture similar to that of Fig~\ref{fig:architectures}(c) to sub-micrometer 
dimensions, and operating it \emph{entirely} on a 1~K refridgeration stage. This would allow fast gate operation times and overcome the problem of limited cooling power, 
typically in the sub-mW range, which dilution-refrigerator based approaces face. Miniaturized linear Paul traps for electrons, with typical 
electron-electrode distances of 500~nm could achieve secular frequencies of $2\,\pi\times20$~GHz and depths of 10~meV, with moderate trapping voltages 
of less than 1~V. The parametric upconversion mechanism, described in Sec.~\ref{sec:hamiltonian}, applied to this case would allow coupling to superconducting 
resonators with frequencies above $2\,\pi\times100$~GHz \cite{Goy1983, Zmuidzinas1994}
enabling ground-state cooling of the electrons in $\sim4$~ns. Electron transport, swapping and entangling gates could be performed in time of order 
0.1~ns. To read out the electron motional state, mapping to a superconducting qubit, as outlined above, is one option, but an alternative option would be
dispersive circuit-CQED type read-out \cite{Blais2004} on the \{$\ket{0},\ket{1}$\} manifold of the electron motion. 
A number of technical challenges would need to be overcome in such an approach. Device miniaturization will not be feasible before the electrode surface noise sources 
are eliminated at cryogenic temperatures, for example reduction by three orders of magnitude over current values would imply electron heating rates of order $3\times10^4$~quanta/s in the example 
mentioned here. In addition, the technology of millimeter wave sources and resonators in the millimeter frequency band, above $2\,\pi\times$100~GHz, 
would need to be adapted to the high-fidelity, low-loss demands of QIP applications.  

In summary, we have proposed a parametric frequency conversion scheme which can couple the motion of trapped particles to solid-state
quantum circuits, and does not rely on non-linear solid-state devices. This scheme allows swapping and entangling operations between electrons and superconducting 
electronics, and can be used to initialize and read-out the state of an electron, as well as to use the electron spin as a quantum memory for superconducting qubits. 
Using current parameters for the device components we find that all basic operations necessary for QIP can be carried out with fidelities close to 99\%. 
We have described applications of this scheme to hybrid quantum architectures in which both trapped electron spins and transmon circuits serve as processing 
qubits. Our toolbox enables a QIP architecture with electrons, similar to the one currently pursued with trapped ions in segmented traps, but having advantages 
in speed and scalability.
\\

%\begin{acknowledgments}
We would like to acknowledge useful discussions with I. Siddiqi, K. Murch and with P. K. Day. 
This research was funded by the ODNI, IARPA, through the ARO grant 30378, by AFOSR  through the ARO grant FA9550-11-1-0318,
by NSF under NSF-DMR-0956064, NSF-CCF-0916303, and by Agilent under ACT-UR 2827. All statements of fact, opinion or conclusions contained herein are 
those of the authors and should not be construed as representing the official views or policies of IARPA, the ODNI, or the U.S. Government.
%\end{acknowledgments}

\appendix

\section{Parametric drive of the electron motion \label{app:param-drive}}

As discussed in the main text, the parametric coupling can be switched on by driving classical electron motion. 
Electron motion can be driven in the $y$ direction, but also in the $x$ direction. To achieve the
latter, we can split the trapping ring electrode into two half rings on the sides of the $yz$ plane, and apply a classical 
out-of-phase drive to the two sides. This option comes at the expense of a factor of 2 reduction 
in the parametric coupling rate and here we focus on driving the $y$ motion. The trap drive and the parametric drive of 
the electron motion are detuned from the superconducting electronics by $\approx$\wy. 
In order to drive electron motion in the $y$ direction at $\Omega_d \approx 2\pi\times6.5$~GHz and $A_d=$\Ad, we apply 
an oscillating voltage of amplitude \Vd~on the ring electrodes labeled $\pm V_{\rm d}$ in Fig.~\ref{fig:AppA} below. Numerical integration of the 
electron equations of motion, with both the trapping potential at $\Omega_{\rm tr}$ and the drive at $\Omega_{\rm d}$, shows that the trap is stable, 
and motional sidebands appear at frequencies $\Omega_{\rm d}+n\Omega_{\rm tr}\pm\omega_{i},\,n=0,\,\pm1,...$. If $\Omega_{\rm tr}=\Omega_{\rm d}$ 
only sidebands at $\Omega_{\rm d}\pm\omega_{i}$ are present, and this can be a preferable configuration. 

  \begin{figure}[!hb]
    \includegraphics[width = 4 cm]{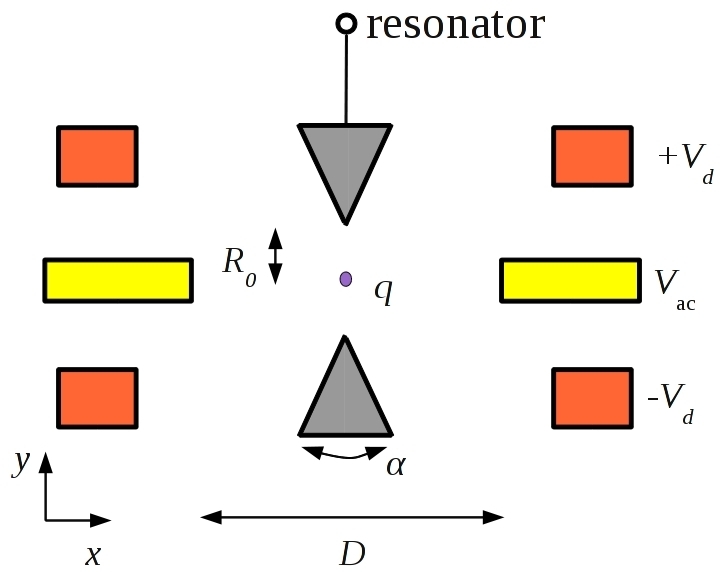}
    \caption{ 
    Cross section of the trap, with the different electrodes labeled. We use the parameters $D=$\D, $R_0=$\R, opening angle
    $\alpha = 20^o$. In order to drive electron motion in the $y$ direction at $\Omega_d \approx 2\pi\times6.5$~GHz and $A_d=$\Ad, 
    we apply to the top and bottom drive electrodes (orange) an oscillating drive with opposite amplitudes $\pm V_{\rm d}\cos(\Omega_{\rm d}t)$, 
    where $V_{\rm d} = $\Vd. 
    \label{fig:AppA}
    }
  \end{figure}

It is interesting to consider the limits of applying the proposed parametric scheme to trapped ions, by analyzing the influence on the trapping pseudopotential
when $\Omega_{\rm tr}=\Omega_{\rm d}$. 
The parametric pump field generates a pseudopotential which is not significant for electrons under 
the trapping conditions we described above. The situation is different for ions, because of their lower secular frequencies. To see this, we compare two 
energy scales: The strength of the pseudopotential, $U_{\rm ps,d}$, which arises from the parametric drive when the driven motion amplitude is $A_{\rm d}$,
and the trapping potential with curvature $\frac{1}{2}m\omega^2$. The ratio of the two is 
$\frac{U_{\rm ps,d}}{m\omega^2 A_{\rm d}^2 }\approx \frac{1}{4}\left(\frac{\Omega_{\rm d}}{\omega}\right)^2$. 
So the pseudopotential arising from the parametric drive scales quadratically with the driven motion amplitude, and with the frequency step-up. For example, 
for $^9$Be$^+$ with secular frequency of 2~MHz in a trap such as the one descirbed here, the limiting frequency for $A_{\rm d} =$\Ad is approximately 
$2\pi\times2$~GHz. For higher frequencies it becomes hard to control non-linearities in the trap potential.

\section{Decoherence of the electron motion \label{app:el-heating}}

To estimate the effect of fluctuating electrical-dipole like noise sources on the trap electrodes, we incoherently sum the contributions of all dipoles on the 
surface of the electrodes. For each one of the conical tips with opening angle $\alpha$, we find that the noise is reduced over the  noise generated by a flat 
surface. For low opening angles the noise level at distance $R_0$ from the tip is well approximated by 
\begin{equation}
  S_{\rm E}^{\rm tip}(R_0,\alpha) \approx (\alpha/10)S_{\rm E}(R_0,\pi)\,,
\end{equation}
where $S_{\rm E}(R_0,\pi)$ is the noise at a distance $R_0$ from a flat surface (i.e. a cone of opening angle $\pi$). Similarly, for a ring electrode similar 
to the one in Fig.~1, the noise contribution is estimated at 
\begin{equation}
 S_{\rm E}^{\rm ring}(D,a) \approx 2\left(1+\frac{2a}{D}\right)S_{\rm E}(D/2,\pi)\,,
\end{equation}
i.e. each one of the top and bottom surfaces of the ring contributes the same noise as a flat plane located a distance $D/2$ from the ion ($S_{\rm E}(D/2,\pi)$), 
and the inside surface of the ring contributes a fraction $2a/D$ of that noise. The two rings which are used to drive the electron motion (orange in 
Fig.~\ref{fig:trap}) can easily be placed a factor of 2 or more further away from the ion compared to the trapping ring electrode, and their contribution can 
thus be neglected. Taking these results into account, and based on the noise value measured in cryogenic traps \cite{Brown2011}, the heating rate for an electron 
trapped at \wy~in the ring trap discussed here, is estimated at \dnh~if the frequency scaling of the noise is $1/f$, and at 
\dnl~if the scaling is $1/f^{3/2}$.

\section{Capacitive coupling of classical signals to the quantum bus \label{app:capacitance}}

In the geometries outlined in Fig.~\ref{fig:trap} and Fig.~\ref{fig:circuit}, the classical drive used to trap the electrons and to pump the parametric action can 
couple to the CPW used as a quantum bus, and cause off-resonant excitations. Conversely, if the CPW couples to the transmission lines used to drive the trap and 
the parametric action, then it will radiatively decay into the transmission lines. To minimize these effects, one needs to capacitively drive opposite ends of 
the $\lambda/2$ CPW resonator (Fig.~\ref{fig:circuit}b) in such a way that the most of the capacitive coupling cancels out, or use some equivalent scheme. 
Capacitive coupling of the CPW to a 50$\,\Omega$ feed line or LC resonator used to drive the trap electrodes will only limit the quality factor at the $10^7$ level 
if the coupling capacitance is limited to below 0.2~fF. Here we describe a scheme which is mainly aimed at cancellation of the off-resonant excitation, while 
achieving far greater reduction of the radiative losses.

To minimize off-resonant excitations, we need to carefully balance capacitances in the device and weakly couple in an additional `fine-tuning' signal. 
One possible solution is outlined in Fig.~\ref{fig:AppC}. The signal, which is capacitively coupled via a parasitic capacitance $C_{\rm p}$, to the 
coupling electrode, is also coupled with an appropriate amplitude 
to the opposite end of the $\lambda/2$ resonator, via the balancing capacitance $C_{\rm b}\approx C_{\rm p}$. Both capacitors are connected to a resonator 
with characteristic impedance $Z = \sqrt{L/C}\approx 50\, \Omega$, and moderate quality factor $Q\approx10^3$, which is used to drive the trap electrodes, and 
helps minimize radiative losses of the $\lambda/2$ resonator. An additional 50~$\Omega$ transmission line is capacitively coupled with $C_{\rm b'}\ll C_{\rm p}$ 
to one end of the $\lambda/2$ resonator, and driven with an adjustable amplitude and phase shift, in order to fine-tune the cancelation of the off-resonant 
excitation. The parasitic capacitances in the ring trap described here are on the order of 0.5~fF, and if they are balanced to 
$C_{\rm p}-C_{\rm b}\approx 10 \,{\rm aF}$, the off-resonant excitation of the $\lambda/2$ resonator will amount to approximately 200 photons. To fine-tune the 
cancellation to the level of $10^{-3}$ photons, the amplitude and phase in an adidtional 50$\,\Omega$ line, coupled by $C_{\rm b'}\approx 10\,{\rm aF}$ needs 
to be adjusted at the 0.4~mV level, provided the phase is controlled to better than 10$^{\rm o}$.

  \begin{figure}[h]
    \includegraphics[width = 4 cm]{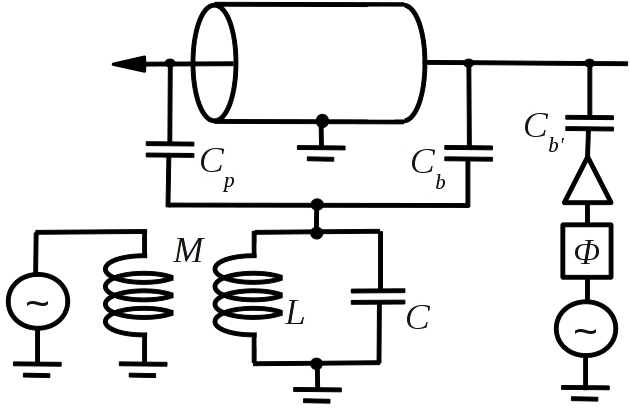}
    \caption{ 
    Circuit to minimize classical pick-up on the CPW quantum bus, and radiative decy of the bus. For simplicity, the ring electrodes 
    (trapping and parametric drive electrodes) are designated by the parasitic capacitances which they contribute. The signal which 
    is capacitively coupled via $C_{\rm p}$ to the coupling electrode, is also coupled with an appropriate amplitude to the opposite 
    end of the $\lambda/2$ resonator, via the balancing capacitance $C_{\rm b}$. An additional $50\,\Omega$ line is directly coupled 
    to one end of the CPW bus via a small coupling capacitance $C_{\rm b'}$.
    }\label{fig:AppC}
  \end{figure}

This configuration also minimizes the inverse effect of radiating from the CPW into the classical-signal transmission lines. Due to the use of an LC resonator 
which is far detuned from the CPW bus and of a weakly coupled transmission line, the radiative loss of the CPW to the external lines will be 
limited to the level of $\kappa<1/{\rm s}$.

\section{Electrical resonator and cavity interaction \label{app:cpw-cavity}}

In order to couple the coplanar waveguide transmission line to the transmon cavity, perhaps the simplest option is for the center conductor 
and one side of the ground plane of the line to extend into the cavity, with an appropriate modification in geometry to maintain the impedance of 
the transmission line constant. 
To estimate the interaction strength of the TEM mode of the CPW to the ${\rm TE}_{011}$ mode of the cavity, we treat the transmission 
line as a collection of electrical dipoles formed between the center conductor and ground. The dipoles arise from the local charge density on the CPW
and they form a continuous distribution over its length. A segment of length $dz$ along the line direction ($z$) has dipole strength 
$\mu(z)\approx\frac{2\,\pi\,d_0\,q_0}{\lambda}\,\sin(2\,\pi\,z/\lambda)$. 
Here $d_0$ is the spacing between the CPW center and signal return conductors $q_{\rm 0} = \sqrt{\frac{\hbar}{2 Z}}$ is the magnitude of charge 
fluctuations in the line, and $\lambda$ is the wavelength of the wave in the CPW. 
If the electric field of the ${\rm TE}_{011}$ cavity mode, $E_{\rm C}(z)$, is aligned with the dipoles (i.e. if it is along the line connecting the center conductor to 
ground), then an upper limit for the coupling strength can be expressed as the integral 
$\hbar G_{\rm lc} = \frac{1}{l}\int_{0}^{l}\mu(z)\,E_{\rm C}(z) dz  = \frac{E_{\rm C,0}\,q_{\rm 0}\,d_0\,l_{\rm eff}}{\lambda}$, 
where $E_{\rm C,0}=\sqrt{\frac{\hbar\omega_{\rm C}}{2\epsilon_0 V}}$ is the magnitude of electric field fluctuation in the cavity, 
and the effective length $l_{\rm eff}$ can be up to order $\lambda/2$. 

We consider a cavity at $2\pi\times$7~GHz, and a CPW with effective impedance of 1~k$\Omega$. A lower limit for $d_0$ is $200 \mu$m, which implies that 
$G_{\rm lc}/\hbar$ can be $2\pi\times$10~MHz, for $l_{\rm eff} = \lambda/2$. Our architecture requires lower values, 
in the 3~MHz range, which can be achieved with appropriate design. 

\section{Electron-transmon quantum electrodynamics \label{app:dressed-transmon}}

The electron-transmon system is at heart a problem of four coupled quantum systems: three oscillators and a qubit. The electron motion, 
intermediate quarter wave resonator, and transmon cavity function as harmonic oscillators, while the transmon acts as a qubit. It is 
illustrative to write the effective four-system problem by an effective Hamiltonian

	\begin{equation}
		H_{\rm eff} = \mathbf{a^\dagger}
		\left( \begin{array}{cccc}
			\omega & g_{\rm p} & 0 & 0 \\
			g_{\rm p} & \omega + \delta &G_{\rm{lc}} & 0  \\
			0 &G_{\rm{lc}} & \omega + \Delta & G_{\rm{tc}}  \\
			0 & 0 & G_{\rm{tc}} & \omega' \\
			\end{array} \right) \mathbf{a}
			= \mathbf{a^\dagger C a},
	\end{equation}
with $\mathbf{a} = (a_x, a_\phi, a_c, \sigma_-)^T$, the vector of excitation annihilation operators for the electron, transmission line, 
transmon cavity, and transmon respectively. For presentation, we have absorbed all the time-dependent factors into the definitions 
of $\mathbf{a}$ and $\mathbf{a^\dagger}$. Such a formulation is useful because the coupling matrix $\mathbf{C}$ contains the relevant 
dynamics. The excitation energies are read off from the diagonal elements, and the coupling rates are read off from the off-diagonal 
elements.

In the limit where the cavity-transmon coupling is the strongest ($G_{\rm tc} \gg G_{\rm lc}, g_{\rm p}$), we can view the eigenstates of the 
cavity-transmon system as the modes of interest, and focus on coupling to the transmon dressed state. 
Then, the problem can be reduced to an effective three-system problem in the following way. First, we diagonalize the cavity-transmon block 
in the limit $\Delta \gg G_{\rm{tc}}$. After the diagonalization we get two vectors: one with a projection mostly onto the transmon 
mode (which we referred to as the 'dressed transmon'), and with a projection onto the cavity mode only of order $G_{\rm{tc}}/\Delta$. 
The second has a projection mostly onto the cavity mode and projects onto the transmon mode also to order $G_{\rm{tc}}/\Delta$.

The first vector represents the operator $\sigma_+ \sigma_- + (G_{\rm{tc}}/\Delta) a^{\dagger}_c \sigma_-$. This is a Hamiltonian 
operator for a dressed transmon mode. The second vector is similar, representing a mode which lives primarily in the cavity. Since we 
have earlier chosen the cavity to be far detuned from the transmon, this mode can be adiabatically eliminated. Removing this dressed 
cavity mode from the basis produces a reduced coupling matrix
	\begin{equation}
		\mathbf{C_{\mathrm{red}}} =
		\left( \begin{array}{ccc}
			\omega & g_{\rm p} & 0  \\
			g_{\rm p} & \omega + \delta &-G_{\rm{lc}} G_{\rm{tc}}/\Delta   \\
			0 &-G_{\rm{lc}} G_{\rm{tc}}/\Delta & \omega \\
			\end{array} \right).
	\end{equation}
	
By adjusting $G_{\rm{lc}}$ and $\Delta$ so that $G_{\rm{lt}} = G_{\rm{lc}}G_{\rm{tc}}/\Delta = g_{\rm p}$, 
we can obtain complete state transfer and entanglement between the electron motion and the dressed transmon, as we discuss in the main 
text.
\\\\

\bibliographystyle{iopart-num}
%\bibliographystyle{unsrt}
%\bibliography{library.bib}
%\bibliography{library}

\begin{thebibliography}{10}
\bibitem{Note}
Here, the terms of odd order in $r_x$ and $r_z$ can be made vanishingly small by symmetry, and the 
higher than quadrartic terms in $r_i$ can be made negligibly small.
\end{thebibliography}

\end{document}